\documentclass[final]{aipproc}
\layoutstyle{6x9}

\begin{document}

\title{Reliability of the Optimized Perturbation Theory for scalar fields at finite temperature}

\author{R. L. S. Farias}{
  address={Departamento de Ci\^encias Naturais, Universidade Federal de S\~ao Jo\~ao del
Rei, 36301-000 S\~ao Jo\~ao del Rei, MG, Brazil} }

\author{D. L. Teixeira Jr.}{
  address={Departamento de Ci\^encias Naturais, Universidade Federal de S\~ao Jo\~ao del
Rei, 36301-000 S\~ao Jo\~ao del Rei, MG, Brazil} }

\author{R. O. Ramos}{
  address={Departamento de
  F\'{\i}sica Te\'orica, Universidade do Estado do Rio de Janeiro, 20550-013
  Rio de Janeiro, RJ, Brazil} }

\begin{abstract}
The thermodynamics of a massless scalar field with a quartic interaction is studied 
up to third order in the Optimized Perturbation Theory (OPT) method. 
A comparison with other nonperturbative approaches is performed such that the 
reliability of OPT is accessed.

\end{abstract}

\maketitle

\vspace{-0.5cm}

Quantum field theory at finite temperature is the natural scenario to deal with phase 
transitions and with thermodynamic properties of equilibrium states. The range of 
applications goes from the understanding of the early universe to the low-energy 
effective theories in particle physics and condensed matter systems. Nevertheless, 
this sort of phenomena face a major complication, the break down of perturbation 
theory, caused by large fluctuations that can emerge in the systems due to infrared 
divergences and close to the critical points. Therefore, nonperturbative methods are 
required in general. In this work we study the thermodynamics of a massless scalar field 
with a quartic interaction in the context of the nonperturbative method of the Optimized 
Perturbation Theory (OPT). We evaluated the free energy up to the to third order in the 
OPT formalism and carried out the calculation of the pressure. The 
reliability of our results for this model is discussed and we also make a comparison 
with another nonperturbative approach, the Screened Perturbation Theory (SPT).



\quad We consider a massless scalar field theory with $g\phi^4$ interaction and with Lagrangian 
density given by
\vspace{-0.4cm}
\begin{equation}
\mathcal{L} = \frac{1}{2}\left(\partial _{\mu }\phi \right) \left( \partial ^{\mu }\phi \right)-
\frac{g^2}{4!}\phi^4.
\label{lagrangian}
\end{equation}
The implementation of the OPT $\mathcal{L}$ is performed through a linear interpolation,
\begin{equation}
\mathcal{L}\to\mathcal{L}^{\delta } = \frac{1}{2}\left( \partial _{\mu }\phi\right)\left( \partial ^{\mu }\phi \right)-
\frac{\delta g^2 }{4!}\phi^{4}-\left( 1-\delta \right) \eta ^{2}\phi ^{2}+\Delta\mathcal{L}^{\delta}_{\mbox{ct}},
\end{equation}
where $\Delta\mathcal{L}^{\delta}_{\mbox{ct}}$ contains the renormalization counterterms required to 
render the theory finite. In $\mathcal{L}^{\delta}$, $\delta$ is a dimensionless bookkeeping parameter 
used only to keep track of the order that the OPT is implemented (it is set equal to one at the end) 
and $\eta$ is a (mass) parameter determined variationally at any giver order of the OPT. We choose 
the variational criterion known as the Principle of Minimal Sensitive (PMS), defined by
\begin{equation}
\frac {d \Phi^{(k)}}{d \eta}\Big |_{\bar \eta, \delta=1} = 0 \;,
\label{pms}
\end{equation}
which is applied to some physical quantity $\Phi^{(k)}$ calculated up to some $k$-order in the OPT.
The optimum value $\bar \eta$ which satisfies Eq.~(\ref{pms}) is a nontrivial function of the couplings 
(and of the original parameters of the theory), leading to the generation of nonperturbative results. 
The interpolation introduces only quadratic terms, which implies that the renormalizability is preserved 
\cite{prdRORmarcus,prd1-lde}.



\quad SPT is simply a reorganization of the perturbative series for thermal field theory \cite{spt}. 
The Lagrangian density in SPT is written as
\begin{eqnarray}
\mathcal{L}_{\mbox{SPT}} &=& -\mathcal{E}_0 + \frac{1}{2}\left( \partial _{\mu }\phi\right)\left( 
\partial^{\mu }\phi \right)-\frac{1}{2}(m^2-m_1^2)\phi^2 
-\frac{g^2}{4!}\phi^4 + \Delta\mathcal{L} + \Delta\mathcal{L}_{\mbox{SPT}},
\end{eqnarray}
where $\mathcal{E}_0$ is the vacuum density term, and a mass term is added and subtracted. Setting 
$\mathcal{E}_0 = 0$ and $m_1^2 = m^2$, the original Lagrangian is recovered. SPT is defined by taking 
$m^2$ to be of order unity and $m_1^2$ to be of order $g^2$, expanding systematically in powers of $g^2$ 
and setting $m_1^2 = m^2$ at the end of the calculation. New ultraviolet divergences are generate in SPT, 
but they can be canceled by the additional counterterms in $\mathcal{L}_{\mbox{SPT}}$.

The mass parameter $m$ in SPT is completely arbitrary. In order to complete a calculation using SPT, a prescription 
for the mass parameter $m$ as a function of $g$ and $T$ is needed. A discussion for different possibilities for this 
prescription is made in \cite{spt-4loop}. We restrict ourselves to the tadpole gap equation,
\begin{equation}
m_t^2 = g^2\frac {\partial \mathcal{F}}{\partial m^2}\Big |_{m_1=m}.
\label{gap-equation}
\end{equation}

\begin{figure}[!htb]
 \centering{
\includegraphics[height=.030\textheight]{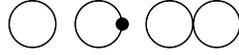}}
 \caption{Contributions to the effective potential at $\mathcal{O}(\delta)$, $F_1$.}
\label{F1}
\end{figure}

\begin{figure}[!htb]
 \centering{
\includegraphics[height=.030\textheight]{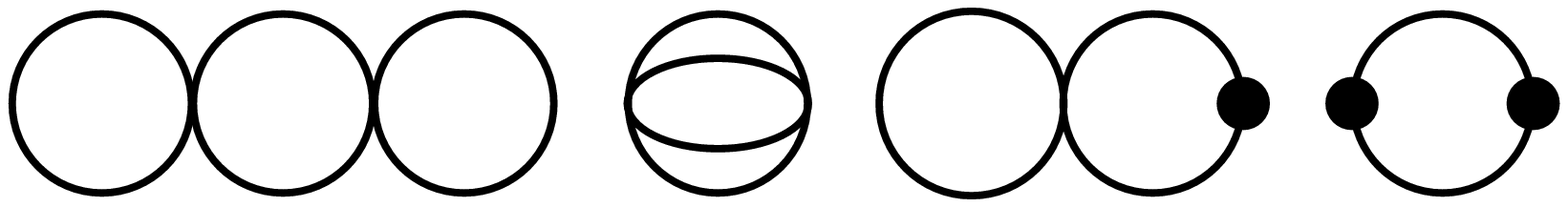}}
 \caption{Contributions to the effective potential at $\mathcal{O}(\delta^2)$, $F_2$.}
\label{F2}
\end{figure}

\begin{figure}[!htb]
 \centering{
\includegraphics[height=.070\textheight]{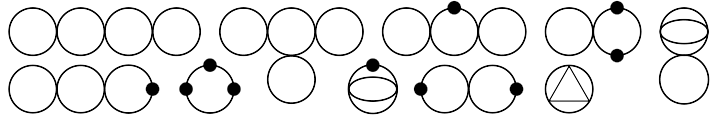}}
 \caption{Contributions to the effective potential at $\mathcal{O}(\delta^3)$, $F_3$.}
\label{F3}
\end{figure}

\quad We apply the PMS, Eq. (\ref{pms}), directly to the free energy, which in terms of 
Feynman diagrams in the OPT formalism at $\mathcal{O}\left(\delta^3\right)$ reads 
\begin{equation}
\mathcal{F} = F_0 + F_1 + F_2 + F_3,
\label{veff-opt}
\end{equation}
where $F_k$ is the free energy evaluated up to order $k$ in OPT and shown in {}Figs. 
\ref{F1}-\ref{F3}. The contributions shown were calculated in the SPT context through 
four-loops expanding in a double power expansion in $m/T$ and $g^2$ and truncating at order 
$g^7$ \cite{spt-4loop}. The expansion required the evaluation of a nontrivial three-loop 
diagram, given by the second term in {}Fig. \ref{F2}, which can be done using the techniques 
developed in \cite{arnold}.

\begin{figure}[!htb]
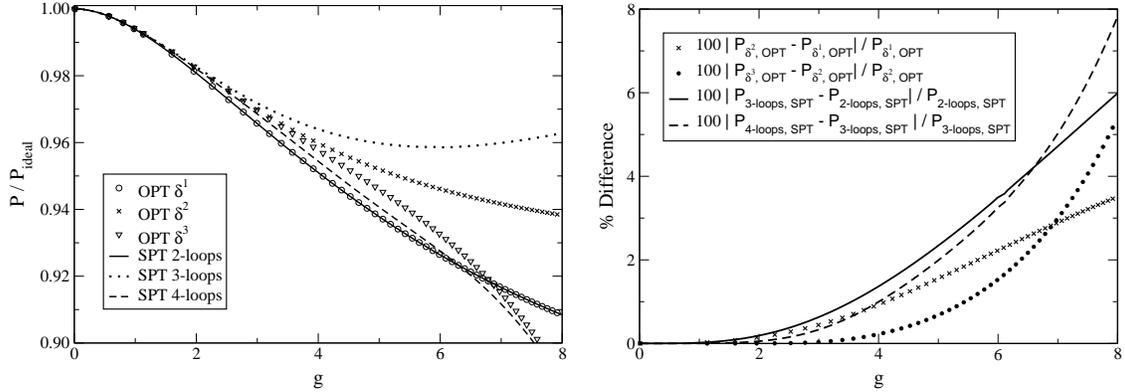

 \centering{
\includegraphics[height=.235\textheight]{graf-pressure.eps} \hspace{0.2cm}
\includegraphics[height=.235\textheight]{perc-diff.eps}}
 \caption{Normalized pressure in OPT and SPT approaches (left) and the percentage 
difference between the results (right).}
\label{graf}
\end{figure}

We are now in position to determine the normalized pressure, which is given by $\mathcal{P} / 
\mathcal{P}_{\mbox{ideal}} = -\mathcal{F}$, where $\mathcal{P}_{\mbox{ideal}} = \pi^2 T^4 / 90$ 
is the pressure of an ideal gas of massless particles. The numerical results obtained 
from the application of the optimization procedure (PMS) in the OPT are shown in Figs. 
\ref{graf}. We compare our results with those obtained from the application 
of the tadpole gap equation at the free energy up to 4-loops in the SPT formalism.

\vspace{0.7cm}

As shown in {}Fig. \ref{graf}, the percentage difference between $P_{\delta^3}$ and 
$P_{\delta^2}$ seems to indicate a convergence in the method faster than SPT. We expect 
to be able, using OPT up to $\mathcal{O}(\delta^3)$, to evaluate thermodynamical quantities 
of interest. Our results suggests a better reliability of the OPT method.

A difference between the methods that appears to be crucial for the improved results 
with OPT is the optimization procedure. In the SPT, the pressure through $k$-loops is 
calculated by using the solution to the $(k-1)$-loop tadpole equation, while in 
the OPT, we use the solution of the PMS applied to the $k$-order free energy to evaluate 
the corresponding $k$-order pressure.


\begin{theacknowledgments}
We thank J. O. Andersen and L. Kyllingstad for providing their data of the 
pressure for comparison. This work was partially supported by CNPq, FAPERJ and FAPEMIG.
\end{theacknowledgments}


\bibliographystyle{aipproc}   
\bibliographystyle{aipprocl}  

\bibliography{PO-Farias}

\end{document}